\documentclass[prl,twocolumn]{revtex4-1}
\usepackage{amsmath} 
\usepackage{graphicx}
\usepackage{color}

\begin{document}

\title{Single active particle engine utilizing a non-reciprocal coupling
between particle's position and self-propulsion}
\author{Grzegorz Szamel}
\affiliation{Department of Chemistry, 
Colorado State University, Fort Collins, CO 80523}

\date{\today}

\begin{abstract}
We recently argued that a self-propelled particle is formally equivalent to a system
consisting of two subsystems coupled by a non-reciprocal interaction
[Phys. Rev. E \textbf{100}, 050603(R) (2019)]. Here we show that this 
non-reciprocal coupling allows to extract useful work from a single self-propelled
particle maintained at constant temperature, by using an aligning interaction to 
influence correlations between the particle's position and self-propulsion. 
\end{abstract} 

\maketitle
\textit{Introduction. --} 
During the last decade, active matter has attracted a lot of interest
\cite{Ramaswamyrev1,Catesrev,Marchettirev1,Bechingerrev,Ramaswamyrev2,Marchettirev2}.
The work in this area started with computer simulations of simple models
and the postulation and analysis of active fluid's hydrodynamics. 
Next, it transitioned to the development 
of statistical mechanical description of model active matter
systems, typically consisting of the so-called self-propelled particles, 
\textit{i.e.} particles that use the energy from their environment to move 
systematically. Very recently, a slightly different class of model systems,
consisting of objects exhibiting non-reciprocal interactions that originate 
from some kind of activity, has become recognized 
for their unusual properties \cite{Brandenbourger,Vitelli2020}. 

Both assemblies of self-propelled particles and
systems exhibiting non-reciprocal interactions 
can be used to construct engines that extract useful work while 
maintained at constant temperature. Proposals in the area of self-propelled
particles include autonomous engines in which active particles power microscopic 
gears \cite{Sokolov,DiLeonardo2010,DiLeonardo2017} 
or push asymmetric obstacles \cite{Pietzonka2019} 
and a cyclic engine that extracts work by changing the properties 
of the confining walls \cite{Ekeh}. Notably, while proposed engines 
work due to the activity, they used its presence somewhat indirectly. 
In contrast, engine protocols that emerged in the area of materials with 
non-reciprocal interactions \cite{Vitelli2020} 
involve direct manipulation of the degree of freedom involved 
in non-reciprocal interactions, which then results in useful work done by
the non-reciprocal force.

We argued recently \cite{Szamel2019} 
that a single self-propelled particle can be viewed as a system
consisting of two subsystems coupled by a non-symmetric (violating Newton's 3rd law), 
\textit{i.e.} non-reciprocal interaction. This suggests 
a possibility to extract useful work from a single self-propelled particle by
controlling directly the degree of freedom involved in the non-reciprocal
interaction, \textit{i.e} the self-propulsion. Specifically, we propose the
following cyclic engine. If the particle's self-propulsion is preferentially 
oriented towards the wall, the wall potential can be relaxed resulting in 
useful work obtained from the particle pushing on the wall. 
Then, if the direction of the self-propulsion can be reversed,
the wall potential can be strengthen back to its original state. While it
seems plausible that useful work would be done by the self-propulsion 
during the relaxing/re-strengthening parts of the cycle, the overall work
balance depends also on the energetic cost of creating specific, atypical states 
of the self-propulsion orientation. Thus, a quantitative analysis of the proposed 
cycle's performance is needed. We use the framework developed recently by
Ekeh \textit{et al.} \cite{Ekeh} and show that useful work can indeed be extracted.

In the following we first discuss a model that can be partially analyzed 
analytically. It consists of a single active particle endowed with self-propulsion 
evolving according to the Ornstein-Uhlenbeck
stochastic process, \textit{i.e.} an active Ornstein-Uhlenbeck particle (AOUP)
\cite{Szamel2014,Maggi2015,Fodor2016}, 
in a harmonic potential. The model used here is different 
from that introduced in Ref. \cite{Szamel2014} in that in addition to the 
self-propulsion there is also a thermal noise and an additional aligning interaction 
that controls the correlations between the self-propulsion and the position of the
particle. For this model it can be shown analytically that useful quasistatic 
work can be extracted from the proposed cycle. 
Next, we use computer simulations to analyze a single active particle with 
self-propulsion of a constant magnitude and a freely-rotating direction, 
\textit{i.e.} an active Brownian particle (ABP) \cite{tenHagen,FilyMarchetti}. 
The particle is moving in two spatial dimensions, under
the influence of a harmonic potential and an aligning interaction controlling 
the correlations between the self-propulsion
direction and the direction away from the center of the potential.
In this case, computer simulation results show that useful work can be extracted 
from the proposed cycle. 

\textit{Simple model engine: an AOUP in a harmonic potential. --}
The equations of motion for the position and the self-propulsion of the particle read,
\begin{align}\label{eomx}
\gamma\dot{x} & = \!\!\!\!\! \underbrace{f}_{\text{non-recip.}} \!\!\!\!\! - k x - 
n\frac{\tau_p}{\gamma} f + \zeta &
\left< \zeta(t) \zeta(t')\right> = 2 \gamma T \delta(t-t')
\\ \label{eomf}
\tau_p \dot{f} & = - f - n x + \eta &
\left< \eta(t) \eta(t')\right> = 2 \gamma T_a \delta(t-t').
\end{align} 
In Eq. (\ref{eomx}) $\gamma$ is the friction coefficient, $x$ is the
position of the particle, $f$ is the self-propulsion and $n$ is the coupling constant
quantifying the strength of an aligning interaction between the position and the 
self-propulsion. The aligning interaction controls the correlations between 
these two quantities. Its specific form was chosen to allow for analytical analysis
of quasistatic cycles. 
Finally, $\zeta$ is the thermal white noise characterized by temperature $T$ 
(note that throughout the paper we use units such that $k_B=1$).
In Eq. (\ref{eomf}) $\tau_p$ is the persistence time of the self-propulsion
and $\eta$ is the noise of the reservoir coupled to the self-propulsion,
characterized by active temperature $T_a$. The first term at the right-hand-side (RHS)
of Eq. (\ref{eomx}) is the non-reciprocal term. This
term induces non-trivial correlations between particle's position
and self-propulsion even in the absence of the aligning interaction 
characterized by coefficient $n$. 

Without the non-reciprocal term in Eq. (\ref{eomx}) and at the temperature equal to
the active temperature, $T=T_a$, the system described by Eqs. (\ref{eomx}-\ref{eomf}) 
is equivalent to two harmonic oscillators coupled by an interaction term
proportional to $n$, in contact with a heat reservoir at temperature $T$. Using 
the Fokker-Planck equation corresponding to Eqs. (\ref{eomx}-\ref{eomf}) one can show 
that the stationary, \textit{i.e.} the equilibrium probability distribution is given by 
$p_\text{eq}(x,f) \propto \exp[-(kx^2/2+n\tau_p fx/\gamma+\tau_p f^2/(2\gamma))/T]$.
In this case, quasistatic cycles defined by changing force constant $k$ and
coupling constant $n$ produce no useful work in the surroundings and 
finite time cycles require external work to be performed on the system. 
It is instructive to note the role of the reciprocal coupling
between position $x$ and self-propulsion $f$: 
for $n>0$, term $n\tau_p fx/\gamma$ induces
negative cross-correlation between $x$ and $f$, $\left<xf\right>_\text{eq}<0$,
where $\left<\ldots\right>_\text{eq}$ denotes averaging over the equilibrium
ensemble.

The presence of the non-reciprocal term in Eq. (\ref{eomx}) changes the system 
in a very profound way. To simplify the analysis we will consider only
the case of the temperature equal to the active temperature, $T=T_a$. 
To analyze the work-dissipated heat balance we use the formalism developed by
Ekeh \textit{et al.} \cite{Ekeh} We recognize that the quantity 
\begin{equation}\label{AOUPutot}
u_\text{tot}(x,f) = 
\frac{1}{2}kx^2+n\frac{\tau_p}{\gamma} fx + \frac{1}{2}\frac{\tau_p}{\gamma} f^2
\end{equation} 
plays the role analogous to the total 
potential energy. This supposition is supported by the analysis of the 
mathematically equivalent system of two overdamped harmonic oscillators
coupled by a reciprocal interaction characterized by $n$ and a non-reciprocal 
interaction that cannot be obtained from a potential. 
We consider a cyclic transformation driven by quasistatic changes of 
force constant $k$ and coupling coefficient $n$. Following Ekeh \textit{et al.}
we can write quasistatic work $W_\text{qs}$ as 
\begin{eqnarray}\label{AOUPWqs}
W_\text{qs} &=& \oint_{\partial\Sigma}
\left[ \left<\frac{\partial u_\text{tot}}{\partial k}\right>_\text{s} dk + 
\left<\frac{\partial u_\text{tot}}{\partial u}\right>_\text{s} du \right]
\\ \nonumber &=&
\pm \int\int_\Sigma \left[\frac{\partial}{\partial n}
\left<\frac{\partial u_\text{tot}}{\partial k}\right>_\text{s}
- \frac{\partial}{\partial k}
\left<\frac{\partial u_\text{tot}}{\partial n}\right>_\text{s}\right] 
dk dn
\end{eqnarray} 
In Eq. (\ref{AOUPWqs}), $\partial \Sigma$ denotes the path in the $(k,n)$ plane
defining the cyclic engine protocol and $\left<\ldots\right>_\text{s}$
denotes averaging over the stationary ensemble, which differs from the 
equilibrium ensemble due to the presence of the non-reciprocal term in Eq. (\ref{eomx}). 
The second line of Eq. (\ref{AOUPWqs}) is obtained using Green's theorem; 
+ and - signs refer to clockwise and counter-clockwise protocols $\partial\Sigma$, and 
$\Sigma$ denotes the part of the $(k,n)$ plane enclosed by $\partial \Sigma$.

For the present model stationary state averages 
$\left<\partial_k u_\text{tot}\right>_\text{s} = 
\left< x^2/2\right>_\text{s}$ and
$\left<\partial_n u_\text{tot}\right>_\text{s} =
\left< \tau_p xf/\gamma\right>_\text{s}$
can be calculated
analytically,
\begin{eqnarray}\label{ux}
\left< \frac{1}{2}x^2\right>_\text{s} = \frac{\tilde{T}}{2\tilde{k}}
\left[1+\frac{\left(\tilde{n}-1\right)\left(\tilde{n}-1+\tilde{n}/\tilde{k}\right)}
{\left(1+\tilde{k}\right)\left(1+\tilde{n}\left(1-\tilde{n}\right)/\tilde{k}\right)}
\right],
\end{eqnarray}
\begin{eqnarray}\label{um}
\left< \frac{\tau_p}{\gamma}xf\right>_\text{s} = 
- \tilde{T} \frac{\tilde{n}-1+\tilde{n}/\tilde{k}}
{\left(1+\tilde{k}\right)\left(1+\tilde{n}\left(1-\tilde{n}\right)/\tilde{k}\right)},
\end{eqnarray}
where $\tilde{T}=T\tau_p/\gamma$, 
$\tilde{k}=k\tau_p/\gamma$ and $\tilde{n}=n\tau_p/\gamma$.
A direct inspection shows that the integrand in the second line of Eq. (\ref{AOUPWqs}) 
$w_\text{qs} = \partial_n
\left<\partial_k u_\text{tot}\right>_\text{s}
- \partial_k
\left<\partial_n u_\text{tot}\right>_\text{s}\neq 0$ and thus
a non-vanishing useful quasistatic work can be extracted from a cyclic process
in the $(k,n)$ plane. 

For illustration, we considered the following cycle in the 
$(k\tau_p/\gamma,n\tau_p/\gamma)$ plane: 
$(1,-.5)\rightarrow (1,.5)\rightarrow (2,.5)\rightarrow (2,-.5)\rightarrow (1,-.5)$.
In the part of the $(k,n)$ plane enclosed by this cycle integrand $w_\text{qs}$ does not
change the sign, which allows us to vary $k$ and $n$ independently \cite{Ekeh}. 
We obtained quasitatic work $W_\text{qs} = 0.545 \, T$. 

Next, we consider finite-time cycles. We note that the average of finite-time work $W$,
\begin{eqnarray}\label{AOUPWt}
W = \int_0^{\tau_c} \left[\frac{\partial u_\text{tot}}{\partial k}\dot{k}
+\frac{\partial u_\text{tot}}{\partial n}\dot{n}\right] dt
\end{eqnarray}
can be expressed in terms of the moments of
the time-dependent probability distribution. Specifically, to evaluate 
$\left< W\right>_t$ we need
$\left< x^2/2\right>_t$ and $\left< \tau_p xf/\gamma\right>_t$.
Here $\left< \dots \right>_t$ denotes averaging over time-dependent 
probability distribution that can be obtained by solving the Fokker-Planck
equation corresponding to equations of motion (\ref{eomx}-\ref{eomf}),
with time-dependent couplings $k(t)$ and $n(t)$.
Since Eqs. (\ref{eomx}-\ref{eomf}) are linear, 
equations of motion for second moments do not involve
higher moments. We obtain the following equations of motion 
for three second moments, 
\begin{eqnarray}\label{eomux}
\frac{\partial}{\partial t}\left< \frac{1}{2}x^2\right>_t \!\!\! &=& \!
-2\frac{k(t)}{\gamma} \left< \frac{1}{2}x^2\right>_t + \frac{T}{\gamma}
\nonumber \\ && \!\!\!\!\!
+ \left(\frac{1}{\tau_p}-\frac{n(t)}{\gamma}\right) 
\left< \frac{\tau_p}{\gamma}xf\right>_t
\\[1ex] \label{eomum}
\frac{\partial}{\partial t} \left< \frac{\tau_p}{\gamma}xf\right>_t \!\!\!
&=& \!
-\left(\frac{k(t)}{\gamma}+\frac{1}{\tau_p}\right)
\left< \frac{\tau_p}{\gamma}xf\right>_t
\\ \nonumber && \!\!\!\!\!
- 2 \frac{n(t)}{\gamma}\left< \frac{1}{2}x^2\right>_t
+ 2\left(\frac{\tau_p}{\gamma^2}
-\frac{n\tau_p^2}{\gamma^3}\right)\left< \frac{1}{2}f^2\right>_t
\\[1ex] \label{eomuf}
\frac{\partial}{\partial t} \left< \frac{1}{2}f^2\right>_t \!\!\! &=& \!  
-\frac{2}{\tau_p}\left< \frac{1}{2}f^2\right>_t + \frac{T\gamma}{\tau_p^2}
\nonumber \\ && \!\!\!\!\!
-\frac{n(t)\gamma}{\tau_p^2}\left< \frac{\tau_p}{\gamma}xf\right>_t .
\end{eqnarray}

\begin{figure}
\includegraphics[width=3.1in]{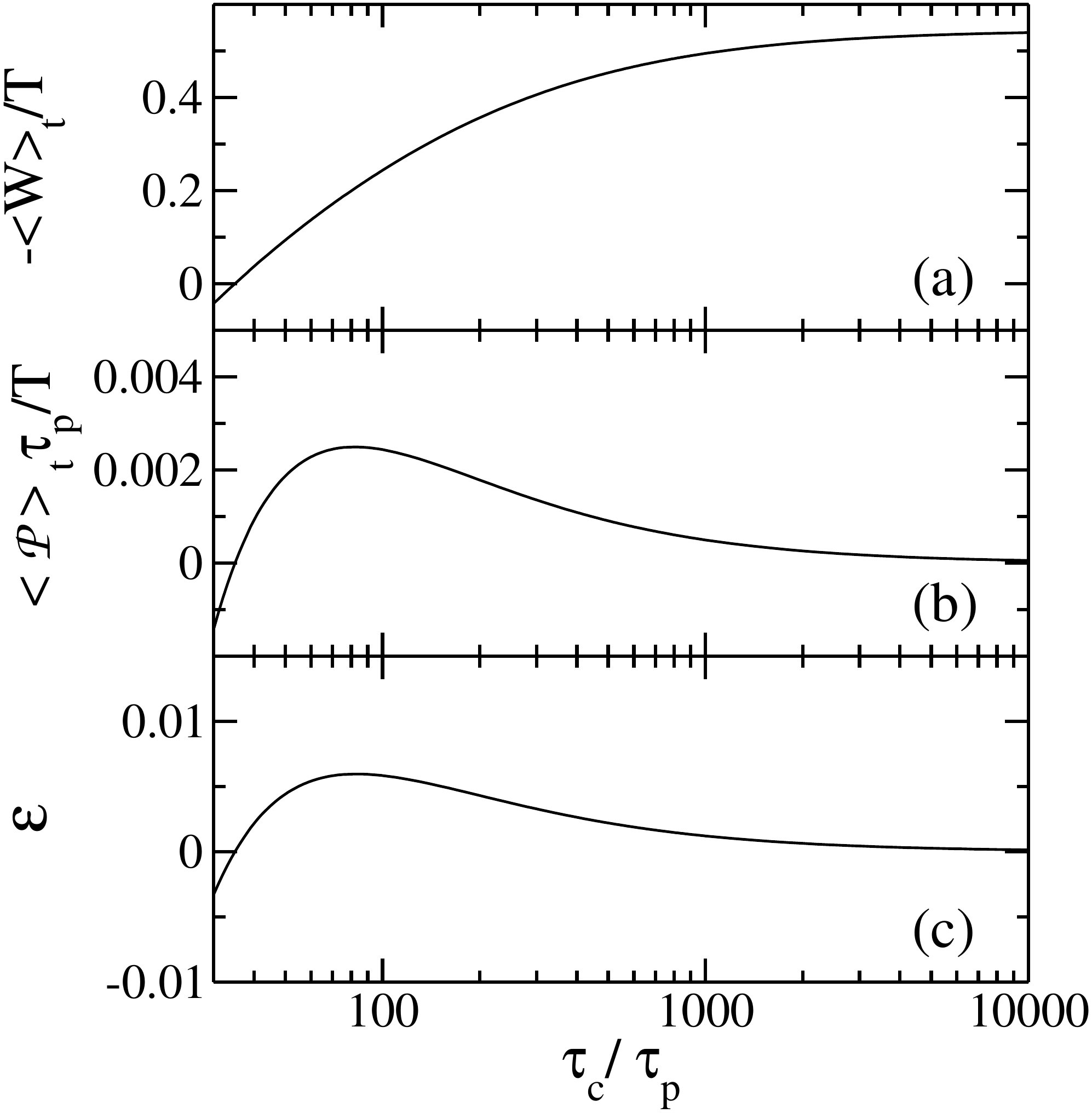}
\caption{\label{fig1} Dependence of the thermodynamic quantities characterizing the 
AOUP engine on the cycle time scaled by the persistence time, $\tau_c/\tau_p$. 
(a) Average useful work normalized by the temperature, $-\left<W\right>_t/T$. 
(b) Average power normalized by the ratio of the temperature and the persistence time, 
$\left<\mathcal{P}\right>_t/(T/\tau_p)$. 
(c) Engine efficiency $\mathcal{E}$.}
\end{figure}

Eqs. (\ref{eomux}-\ref{eomuf}) were solved numerically \cite{numdiff} for the same cycle 
for which we calculated the quasistatic work;  
in the $(k\tau_p/\gamma,n\tau_p/\gamma)$ plane the cycle reads  
$(1,-.5)\rightarrow (1,.5)\rightarrow (2,.5)\rightarrow (2,-.5)\rightarrow (1,-.5)$,
with $k(t)$ and $n(t)$ changing linearly with time in elements 2 and 4 and 1 and 3 
of the cycle, respectively. Cycle time dependence of the resulting average useful work 
(work done in the surroundings) $-\left< W \right>_t$ and 
power $\left<\mathcal{P}\right>_t \equiv -\left< W \right>_t/\tau_c$ 
are shown in Figs. \ref{fig1}(a)-\ref{fig1}(b). We find that the 
results are similar to those of Ekeh \textit{et al.} \cite{Ekeh}: useful work
is extracted only for long enough cycles and power reaches its peak
value for an intermediate cycle time.   

To evaluate efficiency of the cycle we extend the formalism of 
Ekeh \textit{et al.} to the present case of an AOUP engine. 
Briefly, we define heat as work done on the two subsystems by their
respective thermostats,
\begin{eqnarray}\label{AOUPqt}
Q = \int_0^{\tau_c} \left[ \left(\gamma \dot{x} - \zeta \right)\dot{x} +
\frac{\tau_p}{\gamma}\left(\tau_p \dot{f} - \eta \right) \dot{f} \right] dt .
\end{eqnarray}
Again, expression (\ref{AOUPqt}) can be justified by analyzing the mathematically
equivalent system of two coupled harmonic oscillators.
Following Ekeh \textit{et al.} we find that 
\begin{eqnarray}\label{AOUPqw}
\left<Q\right>_t = \left< W \right>_t + \int_0^{\tau_c} f \dot{x} dt, 
\end{eqnarray}
which allows us to calculate the efficiency, 
\begin{eqnarray}\label{eff}
\mathcal{E} = \frac{-\left<W\right>_t}{-\left<W\right>_t + \left<Q\right>_t}
\end{eqnarray}

In Fig. \ref{fig1}(c) we show the cycle time dependence of the efficiency. Again,
our results are qualitatively similar to those of Ekeh \textit{et al.} \cite{Ekeh}: the 
efficiency reaches its peak value for an intermediate cycle time, close to the
cycle time corresponding to the peak value of the power. We note that the 
efficiency of our simple single AOUP engine is several orders of magnitude larger
that that of the engine analyzed in Ref. \cite{Ekeh}.

\textit{An ABP engine. --} The equations of motion for a single active Brownian
particle moving in a two-dimensional harmonic potential and subjected to an 
additional aligning interaction read, 
\begin{align}\label{eomr}
\gamma_t\dot{\mathbf{r}} &= \!\!\! 
\underbrace{\gamma_t v_0 \mathbf{e}}_{\text{non-recip.}} \!\!\! 
- \partial_\mathbf{r} u_\text{tot} + \boldsymbol{\zeta}
& \left<\boldsymbol{\zeta}(t)\boldsymbol{\zeta}(t')\right> =
2 \boldsymbol{I} \gamma_t T \delta(t-t'),
\\ \label{eomp}
\gamma_r\dot{\varphi} &= -\partial_\varphi u_\text{tot} + \eta &
\left<\eta(t)\eta(t')\right> = 2 \gamma_r T \delta(t-t').
\end{align}
In Eqs. (\ref{eomr}-\ref{eomp}) $\gamma_t$ and $\gamma_r$ is the translational
and rotational friction constant, respectively and $\boldsymbol{I}$ is a 
unit two-dimensional tensor. Furthermore, 
$\mathbf{e} \equiv (\cos(\varphi),\sin(\varphi))$ is the orientation vector 
specifying the direction of the self-propulsion and 
$u_\text{tot}$ denotes the total potential energy, which is 
a sum of the harmonic and aligning contributions, 
$u_\text{tot}(\mathbf{r},\varphi) = k \mathbf{r}^2/2 
+ l \hat{\mathbf{r}}\cdot\mathbf{e}$. We note that the coefficients characterizing 
the strength of the aligning interaction for the AOUP and ABP systems have different
dimensions and for this reason we use different symbols to refer to them. 
Finally, $\boldsymbol{\zeta}$ and 
$\eta$ are thermal white noises characterized by temperature $T$. 
We recall that the orientational dynamics of ABPs is often discussed
in terms of rotational diffusion coefficient $D_r=T/\gamma_r$, which physically
corresponds to the inverse of the persistence time of AOUPs.
The first term at the RHS
of Eq. (\ref{eomr}) is the non-reciprocal term. Its magnitude  
is quantified by self-propulsion velocity $v_0$. 
As for the AOUP, the non-

\onecolumngrid

\begin{figure}
\includegraphics[width=6.4in]{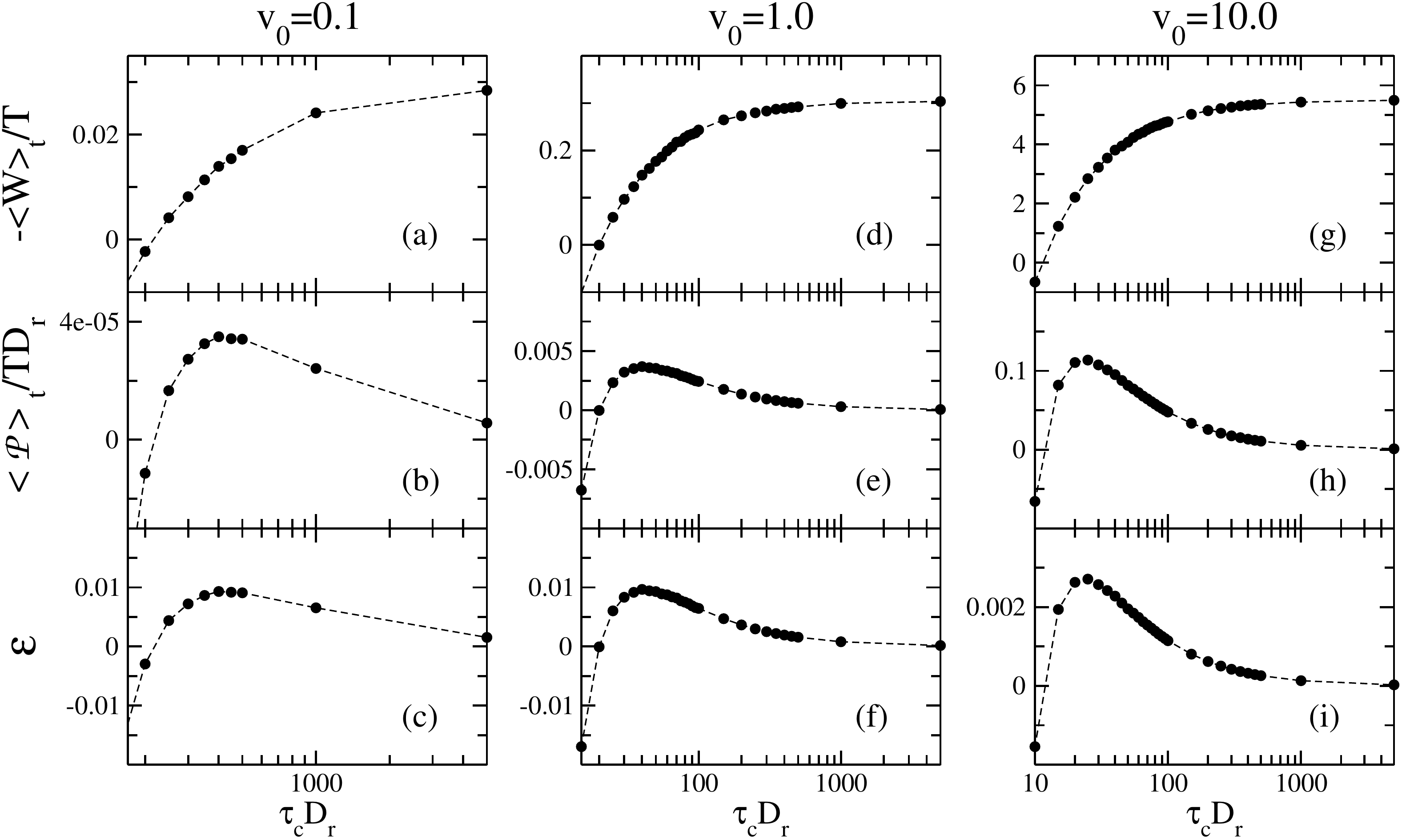}
\caption{\label{fig2} Dependence of the thermodynamic quantities characterizing the 
ABP engine on the cycle time scaled by the rotational diffusion coefficient, 
$\tau_c D_r$, for self-propulsion velocity $v_0=0.1$ (left panels), $v_0=1.0$ 
(middle panels) and $v_0=10.0$ (right panels).
(a,d,g) Average useful work normalized by the temperature, $-\left<W\right>_t/T$. 
(b,e,h) Average power normalized by the product of the temperature and the rotational
diffusion coefficient, 
$\left<\mathcal{P}\right>_t/(T D_r)$. 
(c,f,i) Engine efficiency $\mathcal{E}$.}
\end{figure}

\twocolumngrid

\noindent
reciprocal term induces non-trivial correlations between 
particle's position and orientation, even in the absence of the aligning interaction. 

To analyze the single ABP engine we solved Eqs. (\ref{eomr}-\ref{eomp})
numerically \cite{numsol} with time-dependent couplings $k(t)$ and $l(t)$
following a cycle similar to that used for the single AOUP engine. 
Specifically, in the $(k/(\gamma_t D_r),l/T)$ plane we considered path 
$(1,-1)\rightarrow (1,1)\rightarrow (2,1)\rightarrow (2,-1)\rightarrow (1,-1)$,
with $k(t)$ and $l(t)$ changing linearly with time in elements 2 and 4 and 1 and 3 
of the cycle, respectively. We note that for the single ABP engine the 
magnitude of the self-propulsion velocity is an additional parameter, which
can be varied for a given cycle path.

In Fig. \ref{fig2} we show average useful work $-\left< W \right>_t$, 
power $-\left< W \right>_t/\tau_c$ and efficiency $\mathcal{E}$, calculated
in the same way as for the single AOUP engine,  for a range
of the cycle times and three self-propulsion velocities, $v_0=0.1$, 1.0 and 10. 
Qualitatively, for any $v_0$ the results are similar to those for the 
single AOUP engine. The main trends
are increasing magnitude of both the useful work and the maximum power 
with increasing $v_0$, decreasing efficiency for larger $v_0$ and 
decreasing the cycle time at which the maximum power and efficiency are
attained with increasing $v_0$.

\textit{Discussion. --} We have proposed a general framework for simple
engines utilizing the non-reciprocal interaction that is inherent in many
active particles' models. We have analyzed two examples of such engines, 
that use as working bodies single self-propelled particles. We have shown that
by manipulating correlations between particle's position and self-propulsion one
can extract useful work from monothermal cycles. Generally, these engines 
achieve maximum power and maximum efficiency for finite time cycles. 

We note that our framework can be further simplified: it should be possible to
extract useful work from a single self-propelled particle engine using a 
feedback mechanism based on periodic observation of the self-propulsion direction.
If the direction is pointing towards the wall, the wall potential can be relaxed
and, conversely, if the direction is pointing towards the center of the potential,
the wall potential can be re-strengthen. This new framework
resembles a continuous Maxwell demon proposed, analyzed and experimentally 
realized by Ribezzi-Crivellari and Ritort \cite{Ribezzi}. Its quantitative 
analysis requires taking into account thermodynamic aspects of the information 
content \cite{ParrondoHorowitzSagawa} and is left for a future study.

\begin{acknowledgments}
I thank Elijah Flenner 
for comments on the manuscript. I gratefully acknowledge the support
of NSF Grant No.~CHE 1800282. 
This work was started while I was participating in virtual KITP Program on
the Symmetry, Thermodynamics and Topology in Active Matter; online presentations
at the program are acknowledged as the inspiration. KITP is supported by 
NSF Grant No.~PHY 1748958.
\end{acknowledgments}


\begin{thebibliography}{99}
\bibitem{Ramaswamyrev1} S. Ramaswamy, ``The Mechanics and Statistics
of Active Matter'', Ann. Rev. Condens. Matter Phys. \textbf{1},
323 (2010).

\bibitem{Catesrev} M.E. Cates, 
``Diffusive transport without detailed balance in motile bacteria: does microbiology 
need statistical physics?'', Rep. Prog. Phys. \textbf{75}, 042601 (2012).

\bibitem{Marchettirev1} M.C. Marchetti, J.F. Joanny, S. Ramaswamy, T.B. Liverpool,
J. Prost, M. Rao, and R.A. Simha, 
``Hydrodynamics of soft active matter'', Rev. Mod. Phys. \textbf{85}, 1143 (2013).

\bibitem{Bechingerrev} C. Bechinger, R. Di Leonardo, H. L\"owen, C. Reichhardt,
G. Volpe and G. Volpe, 
``Active particles in complex and crowded environments'', 
Rev. Mod. Phys. \textbf{88}, 045006 (2016).

\bibitem{Ramaswamyrev2} S. Ramaswamy, ``Active matter'', J. Stat. Mech. 054002 (2017).

\bibitem{Marchettirev2} E. Fodor and M.C. Marchetti, 
``The statistical physics of active matter: From self-catalytic
colloids to living cells'', Physica A \textbf{504}, 106 (2018).

\bibitem{Brandenbourger} M. Brandenbourger, X. Locsin, E. Lerner and C. Coulais,
``Non-reciprocal robotic metamaterials'', Nature Communications \textbf{10}, 4607
(2019).

\bibitem{Vitelli2020} C. Scheibner, A. Souslov, D. Banerjee, 
W.T.M. Irvine and V. Vitelli,
``Odd elasticity'', Nature Physics \textbf{16}, 475 (2020).

\bibitem{Sokolov} A. Sokolov, M. M. Apodaca, B. A. Grzybowski, and I.S. Aranson, 
``Swimming bacteria power microscopic gears'', 
Proc. Natl. Acad. Sci. USA \textbf{107}, 969 (2010).

\bibitem{DiLeonardo2010} R. Di Leonardo, L. Angelani, D. Dell'Arciprete, G. Ruocco, 
V. Iebba, S. Schippa, M.P. Conte, F. Mecarini, F. De Angelis, and E. Di Fabrizio, 
``Bacterial ratchet motors'', 
Proc. Natl. Acad. Sci. USA \textbf{107}, 9541 (2010).

\bibitem{DiLeonardo2017} 
G. Vizsnyiczai, G. Frangipane, C. Maggi, F. Saglimbeni, S. Bianchi, and R. Di Leonardo, 
``Light controlled 3D micromotors powered by bacteria'', 
Nature Communications \textbf{8}, 15974 (2017).

\bibitem{Pietzonka2019} 
P. Pietzonka, E. Fodor, C. Lohrmann, M.E. Cates, and U. Seifert, 
``Autonomous Engines Driven by Active Matter: Energetics and Design Principles'', 
Phys. Rev. X \textbf{9}, 041032 (2019).

\bibitem{Ekeh} T. Ekeh, M.E. Cates, and E. Fodor,
``Thermodynamic cycles with active matter'',
Phys. Rev. E \textbf{102}, 010101(R) (2020).

\bibitem{Szamel2019} G. Szamel, 
``Stochastic thermodynamics for self-propelled particles'',
Phys. Rev. E \textbf{100}, 050603(R) (2019).

\bibitem{Szamel2014} 
G. Szamel, 
``A self-propelled particle in an external potential: Existence of an effective 
temperature'',
Phys. Rev. E \textbf{90}, 012111 (2014).

\bibitem{Maggi2015} 
C. Maggi, U.M.B. Marconi, N. Gnan, and R. Di Leonardo,
``Multidimensional stationary probability distribution for interacting active
particles'',
Scientific Reports \textbf{5}, 10742 (2015).

\bibitem{Fodor2016}
E. Fodor, C. Nardini, M. E. Cates, J. Tailleur, P. Visco, and F. van Wijland,
``How Far from Equilibrium Is Active Matter?'',
Phys. Rev. Lett. \textbf{117}, 038103 (2016).

\bibitem{tenHagen} B. ten Hagen, S. van Teeffelen and H. L\"owen,
``Brownian motion of a self-propelled particle'',
J. Phys.: Condens. Matter \textbf{23} 194119 (2011).

\bibitem{FilyMarchetti} Y. Fily and M.C. Marchetti,
``Athermal Phase Separation of Self-Propelled Particles with No Alignment'',
Phys. Rev. Lett. \textbf{108}, 235702 (2012)

\bibitem{numdiff} We used fourth-order Runge-Kutta routine \texttt{rk4}; 
see W.H. Press, S.A.
Teukolsky, W.T. Vetterling and B.P. Flannery, \textit{Numerical Recipes
in FORTRAN}, 2nd ed. (Cambridge University Press, New York, 1992).

\bibitem{numsol} We used the Euler-Maruyama method with time step $10^{-4}$ 
and averaged over at least $5\times 10^4$ realizations. 

\bibitem{Ribezzi} M. Ribezzi-Crivellari and F. Ritort, 
``Large work extraction and the Landauer limit in a continuous Maxwell demon'',
Nature Physics \textbf{15}, 660 (2019).

\bibitem{ParrondoHorowitzSagawa} J.M.R. Parrondo, J.M. Horowitz and T. Sagawa, 
``Thermodynamics of information'', Nature Physics \textbf{11}, 131 (2015).

\end{thebibliography}
\end{document}